# Human Sensing via Passive Spectrum Monitoring

Huaizheng Mu, Liangqi Yuan, *Student Member, IEEE*, Jia Li, *Senior Member, IEEE*

**Abstract** Human sensing is significantly improving our lifestyle in many fields such as elderly healthcare and public safety. Research has demonstrated that human activity can alter the passive radio frequency (PRF) spectrum, which represents the passive reception of RF signals in the surrounding environment without actively transmitting a target signal. This paper proposes a novel passive human sensing method that utilizes PRF spectrum alteration as a biometrics modality for human authentication, localization, and activity recognition. The proposed method uses software-defined radio (SDR) technology to acquire the PRF in the frequency band sensitive to human signature. Additionally, the PRF spectrum signatures are classified and regressed by five machine learning (ML) algorithms based on different human sensing tasks. The proposed Sensing Humans among Passive Radio Frequency (SHAPR) method was tested in several environments and scenarios, including a laboratory, a living room, a classroom, and a vehicle, to verify its extensiveness. The experimental results show that the SHAPR method achieved more than 95% accuracy in the four scenarios for the three human sensing tasks, with a localization error of less than 0.8 m. These results indicate that the SHAPR technique can be considered a new human signature modality with high accuracy, robustness, and general applicability.

## I. Introduction

Human modality recognition is increasingly popular due to the availability of low-cost monitoring and high-resolution sensing devices. Human sensing technology has been widely used in various fields such as indoor navigation, access control, security systems, safety equipment, and autonomous vehicles. For example, first responders can quickly locate injured persons through human sensing technology during rescue operations. In the automotive industry, human sensing applications are mainly used for pedestrian detection and driver/passenger authentication. Biometric attributes, such as face, iris, fingerprints, gait, and voice, are commonly used to achieve human sensing [1-5]. Each sensing modality has its strengths and weaknesses. For instance, facial recognition is convenient and accurate, but its performance is highly dependent on the image quality, and it raises privacy concerns. This paper proposes a novel human sensing modality, Sensing Humans among Passive Radio Frequency (SHAPR) based on the radio frequency (RF) spectrum for human authentication, localization, and activity recognition.

### A. Radio Frequency Spectrum Monitoring

With the rapid development of RF technology, the effective management of spectrum resources has become increasingly vital for efficient information transmission and the utilization of limited spectrum resources. RF spectrum monitoring plays a critical role in this by using specialized equipment and technology to measure the basic parameters and spectrum characteristics of radio emissions [6]. RF monitoring involves obtaining, identifying, measuring, analyzing, and scanning signals using RF observation equipment to limit signal interference and ensure the normal operation of RF equipment. Furthermore, as spectrum monitoring technology advances, it has the potential to be used for human sensing applications.

As an electromagnetic wave, an RF signal can be affected by different materials, such as metal and liquid [7], [8]. Given that a significant portion of the human body is composed of liquid, it can be viewed as an equal volume of liquid that absorbs RF signals present in the environment [9], [10]. The presence of liquid in the human body can affect the RF signal in various ways, including absorption, reflection, and emission, potentially altering the original RF spectrum [11].

### B. Motivation and Proposed Solution

Although RF technology has made significant progress in the field of human sensing, several limitations and challenges hinder the practical application of mainstream RF technology:

1. Active RF sensing technologies may compromise their

Manuscript received June 25, 2023.
This research is supported by the AFOSR grant FA9550-21-1-0224.

H. Mu, L. Yuan, and J. Li are with the Department of Electrical and Computer Engineering, Oakland University, Rochester, MI 48309, USA (e-mail: mu960124@gmail.com; liangqiyuan@oakland.edu; li4@oakland.edu).

main function. For example, WiFi, one of the most popular human sensing technologies, is unsuitable for human sensing applications in a real WiFi communication environment due to unpredictable back-offs and packet sizes [44], [45].

2. Active sensing methods require transmitting and receiving signals actively. While active sensing can detect accurately, it consumes more energy and takes over the limited frequency bandwidth resource. The proposed system allows the selection of sensitive frequency bands, which not only saves energy but also improves the speed of data acquisition.
3. The environment is permeated with a variety of RF signals, yet the impacts and potential harm they pose to the human body remain inconclusive. Hence, passive sensing could reduce the health risk of being radiated by active transmitters. If passive sensing can achieve comparable accuracy to active sensing, passive sensing techniques are superior to active sensing techniques.
4. Mainstream RF technologies such as WiFi, RFID, and mmWave require expensive transmitters, making them cost-prohibitive for many use cases.

At the same time, passive sensing techniques using passive RF (PRF) technology have shown promise for human monitoring applications [14-18]. Unlike active sensing methods, passive sensing does not emit any artificial signals and can utilize the PRF spectrum as a new modality for human signature detection. This approach is also environmentally friendly as it does not utilize the crowded frequency band. Additionally, passive sensing techniques are preferred as they do not pose any health risks from non-ionizing radiation from radio signals. Thus, passive sensing can provide comparable results to active sensing while being a safer and more sustainable solution.

Machine learning (ML) is a popular method to solve classification, regression, and clustering problems. For biometric applications, subject authentication, and activity recognition fall under the classification problem category, similar to other human sensing techniques [12]. The localization task is a regression problem because it classifies a particular region [13]. To implement the feasibility test of the PRF-based human sensing framework, we compared five ML algorithms, selected for their transparency, simplicity, and comprehensibility. Deep learning (DL) neural networks (NN) provide the convenience of automatic feature extraction with weights and biases. Recurrent neural networks (RNN) are one of the most popular NNs used for time-series data processing in recent years. Those ML methods used in human sensing depend on different tasks.

In this paper, we propose a novel human sensing system, SHAPR, emphasizing transmitter-free and device-free solutions. SHAPR system leverages software-defined radio (SDR)-driven PRF sensor-based receivers that passively receive RF signals in various scenarios. By passively receiving RF signals in various scenarios, SHAPR leverages the heterogeneous human RF signature created by the human body's effect on ambient RF signals to enable ML algorithms to efficiently perform classification or regression. We conducted three experiments in four common scenarios, including laboratory, living room, classroom, and vehicle, which involved human authentication, localization, and activity recognition. Our experimental results demonstrate that the proposed SHAPR system achieves 95% accuracy on all three applications in all scenarios, and the average localization error over a sampling distance of 1.8 m is only 0.8 m. This paper's contributions are as follows:

1) We provide an extensive investigation, comparison, discussion, and summarization of the characteristics of active RF and the proposed SHAPR system. Through systematic construction, observation, and experimentation, we confirm the effectiveness and potential of PRF technology.
2) We propose a novel, transmitter-free, and device-free PRF-based human sensing system, SHAPR, which offers a new modality for biometric and human monitoring applications.
3) We evaluate the SHAPR system in three human sensing applications across four experiment scenarios, showcasing its performance, effectiveness, and robustness.
4) We utilize multiple ML techniques to perform classification and regression tasks on processing PRF signals, resulting in an overall classification accuracy greater than 95% and positioning error less than 0.8 m. These results demonstrate the high accuracy, reliability, and general applicability of the proposed SHAPR system.

This paper is structured as follows. Section II reviews the radio spectrum monitoring-related knowledge and applications, existing human sensing techniques, the feature, and applications of SDR, and several ML methods. Section III describes some passive sensing technologies and the features of the SHAPR passive sensing system. Section IV presents the proposed human sensing SHAPR system. Section V shows the experimental results, including authentication, localization, and activity recognition results. Conclusions and future directions are discussed in Section VI.

II. RELATED WORKS

A. *Radio Frequency Technology*

Radio technology can transmit signals through radio waves. The principle of RF technology is based on the theory of electromagnetic waves. The radio signal can be transmitted in an air or vacuum environment. RF information transmission through radio waves requires modulation. When the electric wave propagates in space and reaches the receiver, the radio information can be extracted through demodulation.

Radio has a wide range of applications. The earliest application of radio is the telegraph. The use of the Morse telegraph transmitted information between ships and land stations. With the rapid development of wireless communication technology, the number of available frequency bands has greatly increased. Radio applications are becoming more extensive. Nowadays, the main applications of radio include broadcasting, radar, communication satellites, navigation systems, and networks.

Radio waves are generated by the rapid vibration of the magnetic fields. The speed of vibration is the frequency of the wave, and different frequency bands can be used to transmit various information. The radio spectrum (i.e., frequency bands) comprises electromagnetic waves with frequencies between 3 Hz and 300 GHz. The International Telecommunication Union (ITU) divides the radio spectrum into 12 frequency bands.

The RF spectrum is an essential resource for society. Many countries have rigorous control over the use of the radio frequency spectrum, e.g., the U.S. government owns nearly 60% of the radio spectrum. Due to the limited radio spectrum resources, the pragmatic use of spectrum monitoring becomes crucial. If the RF spectrum use and changes can be recorded and analyzed, the spectrum resources can be effectively utilized. More discussion will be described in Section III.

### B. Software Defined Radio (SDR)

SDR is a standardized modular hardware platform used to monitor the RF spectrum. SDR can realize many operations through software, such as selecting and controlling frequency bands, modem types, data formats, encryption modes, communication protocols, etc. The development of different software modules can achieve different functions, and the software can be upgraded and updated. For example, SDR produces various modulation waveforms and communication protocols and communicates with various traditional radio stations, which broadens the application environment and saves costs.

Many applications of SDR exist for human biometrics. For instance, the SDR is tuned to scan only the frequency bands that are sensitive to human occupancy to improve power efficiency [14]. SDR has been widely used in communications, spectrum monitoring, and RF transmitter identification [19], specifically in improving the power amplifier system and transmitter architecture [20]. SDR can be utilized to realize real-time communication [21], receive the animals' nerve signals [22], recognize gestures through Wi-Fi signals [24], and among others. The position of the mobile station can be estimated by using the signal strength received by the SDR [23]. SDR is used to scan the RF signal spectrums to detect and estimate human occupancy. Moreover, the SDR devices used in our experiments are low-cost, compact, and easy to deploy. In our work, the RTL-SDR with an RTL2832U chip is used to scan and monitor the radio frequencies and collect the raw data. The frequency band that is sensitive to human occupancy is selected through DL methods for further scanning. The specific frequency bands ranging from 300 MHz to 420 MHz are monitored by developing the SDR software.

### C. Machine Learning Algorithms

Traditional ML methods, such as Decision Tree (DT), Support Vector Machine (SVM), k-Nearest Neighbors (k-NN), and Random Forest (RFR), and DL methods such as Convolutional Neural Network (CNN) and Recurrent Neural Network (RNN) have been utilized to sense the human subjects in previous work. DT refines a feature set through analysis of meaningful features based on the domain of interest, such as for human movements [25]. SVM is a statistical method to align features to categories. For example, gender identification of human faces has general features for normal categories of SVM [26]. RFR is also a popular solution to realize human activity recognition with good accuracy [27]. A gait recognition method can uniquely identify humans based on the RFR [28].

Gaussian Process Regression (GPR), an ML model for solving regression problems, is a non-parametric model that uses Gaussian process priors to perform regression analysis on data. GPR can provide the posterior of the prediction result, and when the likelihood is normally distributed, the posterior has an analytical form. GPR has been applied in the fields of image processing and automatic control [29]. GPR is found to be very suitable for solving positioning problems [30], [31] and the prediction results obtained by GPR are highly accurate [32]. The advantages of GPR include using only a few training data points for regression to acquire all position results, predicting high-dimensional data, and flexibly using different kernel functions to construct the relationship between the independent variables and the dependent variables [33]. In SHAPR, the independent variables are PRF spectrums, and the dependent variables are humans occupying positions. GPR model is used to infer the relationship between the PRF spectrums and human occupying positions.

### III. UNDERSTANDING OF PASSIVE SENSING

Radio signals can be classified as active or passive. Active RF involves transmitting a signal from a known transmitter, which is then received and measured by devices using a conversion circuit to detect any changes in the signal. In contrast, passive RF relies on receiving signals from multiple transmitters, with a selected receiver detecting changes in the external measured signal, which is then monitored by a sensitive element and conversion circuit. Passive signals are used in applications such as passive radar and passive RFID technology. Human sensing is achieved by observing specific human characteristics and properties.

### A. Passive Radar

Radar systems can be broadly categorized into active and passive types. Active radar is a traditional system that emits electromagnetic waves to detect, locate, and track targets. In contrast, passive radar (also called bistatic radar) operates by passively observing external radiation sources without emitting energy itself. It detects targets by receiving microwave energy reflected by warm objects or other sources. Passive radar systems consist of an antenna and a susceptible receiving device. Passive radar systems can achieve rapid detection of targets [34]. Passive radar adds sub-array to enhance anti-jamming capability and realizes multi-target positioning [35]. Passive radars utilize signal transmissions to achieve target localization and navigation [36]. Some searchers use satellite illumination to realize multi-band passive radar imaging [37]. However, due to the reliance on third-party transmitters, the operator cannot actively control the transmitter. When the effective radiation power of transmitters is low, the signal between the target and the receiver is blocked, or the signal between the receiver and the transmitter is blocked, and the passive radar cannot be used to validate the signal.

*B. Passive RFID*

RFID is a communication technology that can identify specific targets and read/write related data through radio signals without establishing mechanical or optical contact between the identification system and specific targets. RFID includes a passive or an active tag. The principle of passive RFID is that after the tag enters the reader area, the tag can accept the microwave signal transmitted by the RF identification reader. Relying on the energy obtained by the induced current, the product information stored in the tag can be read. RFID has a wide range of applications. Typical applications include animal chips, car chip anti-theft devices, access control, parking lot control, and automated production lines. The main limitation of RFID technology is that RFID electronic tag information is easily read and maliciously revised. There are many RFID developments such as UHF-RFID passive tags to locate drone positions [38]. Furthermore, RFID technology also can be used for train localization based on passive tags [39]. Some researchers used an RFID tag, which is fabricated from electro-textile materials, and integrated it into clothing to detect body movement for human-technology interaction [40]. RFID as exploited by a CNN supported gate control [41]. RFID technology can also be applied to monitor and control the posture of the cane to help vision-impaired people [42].

*C. The SHAPR System*

Human sensing systems have a broad range of applications, including traditional methods of human authentication, such as facial or fingerprint recognition. These methods require the human subject to touch or be close to the receiver, and the accuracy of the authentication is dependent on the quality of the image. The SHAPR system, however, scans the surrounding PRF spectrum and does not require specific imaging or receiver quality. Additionally, unlike traditional localization methods, the proposed human localization does not require the mapping of radio signal strength at each location. Instead, several SDRs are placed at fixed locations to acquire the PRF spectrum, and human subjects do not need to wear any receiver or tags. Machine learning algorithms are then applied to classify the spectrum alterations and realize the localization. The SHAPR solution also benefits activity recognition, as human subjects do not need to carry multiple sensors to detect their posture.

In this research, we utilized the unique characteristics of passive RF signals and ML methods to achieve human sensing. Passive signals offer several advantages overactive signals.

1. Passive signals do not require a specific signal source or transmitter, so the SHAPR system does not need to allocate any spectrum resources.
2. Passive signals are safer than active signals as they do not emit any radiation and rely on receiving signals passively, leading to lower power requirements.
3. The proposed SHAPR system is realized by acquiring the surrounding passive RF signal. Any signal even noise can be used in passive sensing.
4. Passive signals are more friendly than active signals because passive signals are relatively easy to deploy and have low maintenance costs. Based on the above-mentioned advantages, human sensing applications realized by using passive signals will significantly broaden the application of the RF spectrum.

IV. METHODOLOGIES

The proposed SHAPR system mainly includes three parts, (1) RTL-SDR sensor hardware components, (2) data preprocessing, and (3) signal exploitation with MLs including a regressor and classifiers. In the PRF sensor part, the PRF sensors are used to collect raw data in different environments for specific human sensing applications. Data collection in the most sensitive frequency band of the human body is highlighted in this part. In the data preprocessing part, we analyze the raw in-phase and quadrature (I/Q) components and convert to average power, which is used as the feature vector in the spectrum data. In the signal exploitation part, five classifiers and one regressor are proposed to extract, fit, and predict the dataset.

*A. The SHAPR System Structure*

This paper presents passive human sensing with radio frequency classification approaches utilizing a new biometric sensing modality. A passive spectrum monitoring method is developed to collect human subject signature samples, which are used to train ML algorithms to achieve human sensing.

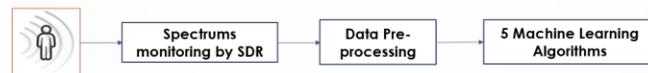

Fig. 1. Flowchart of the proposed SHAPR system.

The proposed SHAPR system is shown in Fig. 1. Part of the indoor RF signals reach the SDR antenna after passing through the human body. SDR passively receives RF signals at a customer frequency band, including signals passing through the human body and other cluttered invalid signals. Different subjects performing tasks in different scenarios lead to different backscattering, refraction, and absorption of signals passing through the human body, which leads to different spectrum signatures received by SDR. The collected dataset trains ML models to recognize different subjects, activities, and positions, which enables human sensing.

RTL-SDR is a software-defined radio (SDR) based on digital video broadcast technology (DVB-T) television (TV) tuners with RTL2832U chips, as shown in Fig. 2, which is used to collect passive RF signals in our research. SDR uses modern software to control the traditional wireless communication technology of hardware circuits. The value of software radio technology is that the realization of the communication functions of traditional radio communication equipment can be developed by software versus designing new hardware.

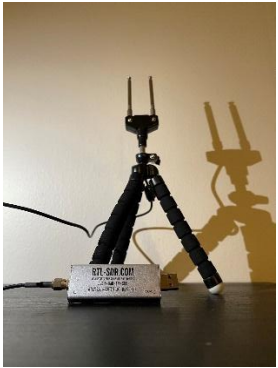

Fig. 2. SDR Antenna and RTL2832U chips.

RTL-SDR collects channel state information (CSI), but CSI is not suitable for passive signals. The main reason is that the passive signal originates from an uncertain direction, which causes uncertainty in the angle information contained in the CSI information which limits the ML model construction. The second reason is that the influence of the human body on the PRF signal is generated in a frequency band rather than a specific frequency and using CSI will increase the difficulty of ML fitting. Therefore, we only use received signal strength indication (RSSI) as the feature of the human spectrum. The collected data is the surrounding RF signals from the scenarios. Several RTL-SDRs are utilized to collect data, which can be conducted to scan the frequency bands from 24MHz to 1760MHz. Our previous research shows that some of the sensitive frequency bands for human occupancy are from 320 to 420MHz. In this paper, 300MHz to 420MHz is selected as our experimental setup. For each frequency, we collected $N$ CSI samples to calculate the RSSI average power in dB, which should be calculated as follow:

$$P(f) = 10 \cdot log_{10} \frac{\sum_{i=1}^{N}\left(\frac{s_f(i)}{127.5} - 1\right)^2}{N/2}, \quad (3.1)$$

where signal average power $P$ is a function of the frequency band center $f$. $N$ is the number of samples in each frequency band, which is 4800 in our experiment. $s_f(i)$ is the value of the $i$-th raw data received by the SDR device when the frequency band center is $f$.

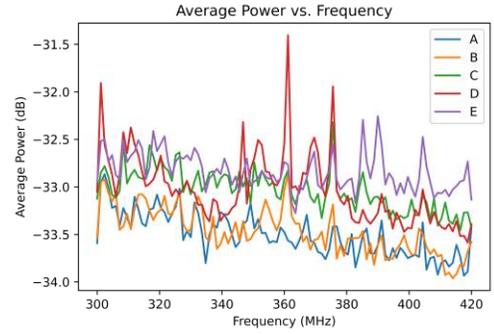

Fig. 3. The average power curve from 300MHz to 420MHz for five SDR antennas at different locations.

Fig. 3 shows an example of the average spectrum power collected by SDR. SDR antennas at different locations are labeled as A, B, C, D, and E. There are great differences in the PRF signals collected by the SDR antennas at different locations, which means that the PRF signals have different characteristics in different locations. The possible reasons for these different characteristics are the propagation direction of the RF signal in the scenario, the nonlinear propagation, the distance between the human body and the SDR, the influence of other objects, etc.

*B. The SHAPR System Mechanism*

For the SHAPR system, we do not know a priori information about the transmitters present in the scenario. Therefore, the mechanisms that enable human sensing can be viewed as the superposition of multiple ambient signals on the human body, including multipath absorption, reflection, refraction, scattering, and so on. Also, it is an open question whether the human body transmits RF signals or not. We analyze the human sensing mechanism based on three human monitoring applications.

**Authentication**. Five ML algorithms are used to associate spectrum samples with different human subjects, including DT, SVM, k-NN, RFR, and RNN. The input of the five ML classifiers is the average power spectrum density, and the output is the index of the human subjects. A total of 140 samples of the power spectrum density are split into 70% training data set and 30% testing data set. As highlighted, since the subjects are known, the scenario represents the opportunity for human subject authentication. Inherent in the analysis is that movement supports detection and recognition, and SHAPR classifies the subjects. With a known training signature of specific people, the spectrum can be used to not only classify a human subject; but as compared to a known signature, ID a specific person who has appropriate access assuming they consented to the P-RF collection to create their profile – as consistent with fingerprint access.

**Localization**. Two localization experiments are conducted in a mobile vehicle and static classroom. The vehicle used in our experiment has four seats. One human subject sat at each seat in order. Each seat has an index. Five

SDRs are used to collect raw data. And then, five ML models, SVM, k-NN, DT, RFR, and RNN, are trained to classify different human subjects' occupancy locations. The input is the normalized spectrum density, and the output is the index of human location. In the classroom, a human subject occupies one location when five SDRs scan the spectrum simultaneously. Only one human subject is inside the environment during the experiments.

The coordinate-level localization method is conducted with GPR. The first is to build a training dataset, where the input is the spectrum power collected by five SDRs, and the output is the human occupancy coordinates. The second is to set a mean and kernel function. The kernel function represents the covariance function. The third is to combine the covariance function and training set to calculate a covariance matrix. The fourth is to provide a predicted dataset and put the input of the predicted spectrum power into the trained model. The final is to get the prediction position coordinates by the means, covariance matrix, and new dataset.

The details of how SHAPR uses GPR are described in this Section. The spectrum density of each known location and corresponding coordinates are the training set $D$ for GPR

$$D = \{x_i, y_i\}_{i=1,\ldots n}. \quad (3.2)$$

The inputs vector $x$ represents the stacked spectrum of multi-sensors. The output $y$ is the coordinates of the human subject location. $i$ represents the sample index in $D$. The collected sample amount is denoted by $n$. The mean $\mu(x_i)$ and covariance $k(x_i, x_j)_{i,j=1,\ldots n}$ are determined as the following:

$$\mu(x_i) = E[y_i], \quad (3.3)$$

$$k(x_i, x_j)_{i,j=1,\ldots n} = \sigma^2 \exp\left(-\frac{\|x_i - x_j\|^2}{2l^2}\right), \quad (3.4)$$

$$y_i \sim GP\left(\mu(x_i), k(x_i, x_j)\right), \quad (3.5)$$

where the mean $\mu(x_i)$ is equivalent to 0 in Gaussian process. $\sigma$ and $l$ are the hyperparameters representing the variance and length scale respectively, which can be acquired by maximum likelihood estimation (MLE). The joint prior distribution of predicted and training values can be obtained as

$$K = \begin{bmatrix} k(x_1, x_1) & \cdots & k(x_1, x_n) \\ \vdots & \ddots & \vdots \\ k(x_n, x_1) & \cdots & k(x_n, x_n) \end{bmatrix}, \quad (3.6)$$

where each row can be defined as

$$K_* = [k(x_*, x_1) \ldots k(x_*, x_n)], \quad (3.7)$$

and the diagonal elements can be expressed as

$$K_{**} = k(x_*, x_*), \quad (3.8)$$

resulting in

$$\begin{bmatrix} y \\ y_* \end{bmatrix} \sim N\left(0, \begin{bmatrix} K & K_*^T \\ K_* & K_{**} \end{bmatrix}\right), \quad (3.9)$$

where $x_*$ and $y_*$ represent the spectrum and coordinates in the testing dataset, respectively. The expected value of the predicted position can be determined by

$$\hat{y}_* = K_*^T K^{-1} y. \quad (3.10)$$

Thus, the coordinates of the predicted coordinate can be calculated by the trained model. Euclidean distance $d$, i.e., root mean square error (RMSE), between the actual and predicted positions is used to get the prediction error as

$$d(\hat{y}_*, y_*) = \sqrt{\sum_{i=1}^{n}(\hat{y}_* - y_*)^2}. \quad (3.11)$$

**Activity Recognition**. Human subjects produce different RF characteristics based on known activities. These activity characteristics result from dynamic and static differences in human activity, differences in human posture, and differences in position in a scenario. Eight typical activities, which include using Smartphone, Sitting, Watching TV, Walking, Standing, Exercise, writing on the Board, and simulating Falls; are recorded for ML training and classification testing.

V. EXPERIMENT AND RESULTS

In our experiments, the SHAPR system is validated on four scenarios, i.e., laboratory, living room, classroom, and vehicle, and three classification tasks, i.e., human authentication, grid localization, and activity recognition and a regression task, i.e., coordinate localization. In this section, multiple ML algorithms are used to demonstrate the high accuracy, applicability, and generality of the proposed system.

*A. Experimental Setups*

To eliminate the interference from artificial signal transmissions such as broadcast signals, TV channels, Wi-Fi, laptops, mobile phones, etc.; the frequency bands used in the SHAPR system evaluation are from 300MHz to 420MHz, with the step size of 1.2 MHz. There are five SDRs in our experimental setting, so each sample has 505 features. The parameter settings for the SDR device are listed in Table I.

TABLE I
SDRs SETUP IN EXPERIMENTS.

| Name of Devices | RTL-SDR with RTL2832U chip |
|---|---|
| Frequency Range | 300 MHz – 420 MHz |
| Scanning Step | 1.2 MHz |
| Sampling Rate | 2.4 MHz |
| Duration | 2 milliseconds per frequency band |
| Feature Number | 505 |

In general, metal objects, or liquids have a certain effect on the transmission of the signal. Considering the impact of diverse environments, the proposed SHAPR human sensing system is tested and verified on several scenarios for each task. These scenarios and tasks are set based on previous literature and daily life experience.

Human authentication and activity recognition task settings include the laboratory scene due to its relatively narrow space and the characteristics of multiple sources of interference.

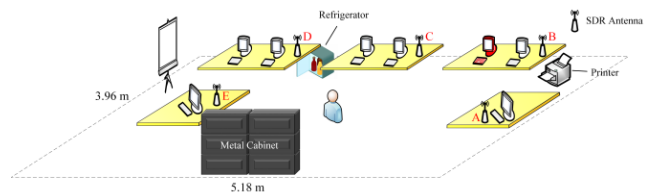
Fig. 4. Illustration of the laboratory scenario is used for human authentication and activity recognition.

As shown in Fig. 4, there are multiple computers, a refrigerator, a printer, and other electronic equipment in the laboratory, which have potential for PRF interference. Five SDR antennas are placed on the table. Subjects may be stationary or moving in the laboratory according to experimental instructions. The red computer is connected to the SDRs to collect the PRF signal. In particular, the presence of the metal cabinet in the laboratory is also a challenge due to the metal-sensitive property of PRF signals.

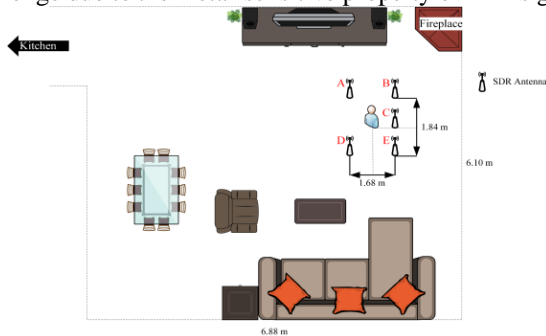
Fig. 5. Illustration of the living room scenario is used for human authentication.

Fig. 5 shows the living room scenario. Five SDR antennas are placed on the ground and surrounded the human subject. Compared to the lab scenario, the living room scenario has a larger range but less electrical interference. The electronic interference in the living room mainly comes from the TV, and other non-metallic furniture such as sofas, tables, and chairs have minimal interference to the PRF signal. Human authentication in the living room scenario is a complementary experiment relative to the laboratory. Compared to activity recognition, human authentication is a more challenging task. Since the human body has different variables such as position, shape, orientation, height, etc. Human authentication requires subjects to be fixed in the same position and maintain the same posture so that the influence of other variables is reduced to the minimum. Therefore, we again conduct secondary experiments in the living room scenario to improve the credibility of our proposed PRF human sensing system on the human authentication task.

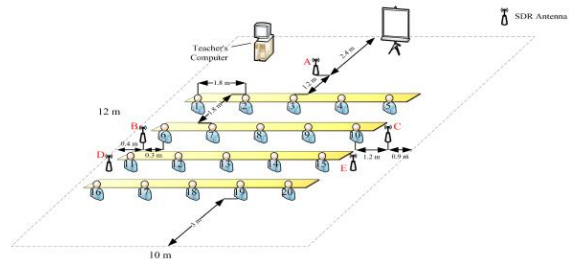
Fig. 6. Illustration of the classroom scenario is used for human localization.

Fig. 6 shows the classroom scenario with the largest area. Five SDR antennas are placed on the ground. Subjects conduct data collection at twenty locations. Compared with the laboratory and living room, the classroom has the largest area, so human localization in the classroom scenario is also a challenging task.

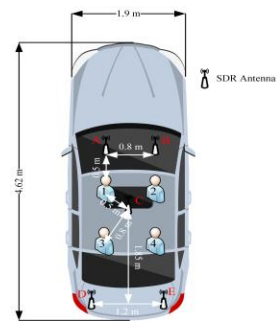
Fig. 7. Illustration of the vehicle scenario is used for human localization.

Human localization in vehicle scenarios is an interesting task. Subjects collected data in the driver, front passenger, and two rear seats. Since the vehicle scenario is surrounded by metal, the RF signal from the outside cannot penetrate, and the position of the subject is closer, which leads to difficulties in human localization. Table II summarizes the data collection in these scenarios including the classification category amount and the sample amount in each category.

TABLE II
DATA COLLECTION IN EXPERIMENTS.

| Task | Scenario | Category Amount | Sample Amount (Per category) |
|---|---|---|---|
| Authentication | Living Room | 7 | 20 |
|  | Laboratory | 12 | 100 |
| Localization | Vehicle | 4 | 20 |
|  | Classroom | 20 | 20 |
| Activity Recognition | Laboratory | 8 | 100 |

*B. Authentication*

The authentication task is conducted in two scenarios, including the laboratory and the living room, demonstrated in Fig. 4 and Fig. 5, respectively. A group of unoccupied samples is collected as a comparison. ML algorithms are used to realize authentication. The authentication results of the DT, SVM, k-NN, RFR, and RNN are shown in Table III.

TABLE III
HUMAN AUTHENTICATION ACCURACY FOR DIFFERENT ML ALGORITHMS IN THE LABORATORY AND THE LIVING ROOM SCENARIOS.

|  | DT | SVM | *k*-NN | RFR | RNN |
|---|---|---|---|---|---|
| **Laboratory** | 84.4% | 76.4% | 85.9% | **95.6%** | 89.0% |
| **Living Room** | 90.5% | 97.6% | 97.6% | **98.7%** | 92.9% |

Table 3 shows that authentication of human subjects can be realized with 95.6% and 98.7% in the laboratory and living room, respectively using the random forest regression (RFR) method. In both scenarios, the RFR algorithm achieves the best accuracy in the authentication task; while it is difficult to discern the relative importance of the other methods and hence, all should be considered for system implementation. The experimental results imply that different human subjects can produce different signatures on the PRF spectrum. The experiment requires the subjects to remain static at the indicated location for data collection, so the subject's behavioral habits do not have an impact on the PRF spectrum. Thus, the reason for the success in the authentication task may be the difference in the subjects' physiological information, such as height, weight, body type, age, gender, body water ratio, etc. Fig. 8 shows the confusion matrix for the human authentication task in the lab and in the living room.

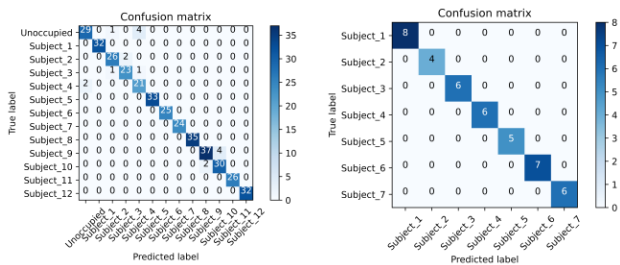

(a) Laboratory  (b) Living Room
Fig. 8. The confusion matrix of authentication in (a) laboratory and (b) living room scenario by using RFR.

In particular, the authentication task is compared in two different scenarios. The living room is relatively more straightforward than the laboratory because there are some electronic devices and mental objects in the laboratory that may interfere with the spectrum. Also, different subjects are utilized for the experiment in the laboratory and the living room, i.e., the experimenters and family members. We obtained more subjects for the experiments in the laboratory because the laboratory has more experimenters available than in the living room environment, which is in line with the actual use case. Also, family members (living room) had a more significant variance in age and physiological status compared to experimenters who are all students (laboratory). From the results of Table 3 and Fig. 8, the reason for the overall higher accuracy of the living room may be the physiological difference between the subjects. Regardless, the RFR accuracy is greater than 95% in both the laboratory and the living room, which demonstrates the capability of the proposed SHAPR system for human authentication.

*C. Localization*

The localization task is also developed in two scenarios, including the classroom and the vehicle. For the vehicle scenario, the vehicle space is smaller than the classroom. In the vehicle scenario, the positions of the subjects are fixed: driver, front passenger, and two rear seats, each indexed with 1, 2, 3, and 4. However, in the classroom scenario, a simple detection task is not enough because of the large area and the uncertainty of the subject's location, the specific coordinates of the subject need to be determined. In the vehicle scenario, the four locations are classified as gridded locations, while in the classroom, we use the data collected at the gridded locations for regression. The classification result on the gridded location at the vehicle scenario is shown in Table IV.

TABLE IV
HUMAN LOCALIZATION ACCURACY FOR DIFFERENT ML MODELS ON THE GRIDDED LOCATION.

|  | DT | SVM | *k*-NN | RFR | RNN |
|---|---|---|---|---|---|
| Vehicle | 91.6% | **99.1%** | 95.8% | **99.1%** | 95.8% |

Among the localization solutions in the vehicle scenario, both SVM and RFR achieved the highest accuracy with 99.1%. These results infer that the human subject can be localized on a grid by using PRF. From our experimental experience, the location of the human signature can make a big difference in the PRF spectrum. Compared with the authentication and activity recognition tasks, the subject's distance from the SDR location has a more intuitive representation on the PRF spectrum. Five SDR receivers are used in the experiment, which can effectively eliminate the influence of RSSI that does not contain phase information. Therefore, obtaining very high accuracy in localization tasks is our expectation. However, the low accuracy of the DT classifier on the localization task gets our attention. The confusion matrixes of the DT are shown in 0, which shows that only position 1 is mislabeled between the front and back passenger (Fig. 7).

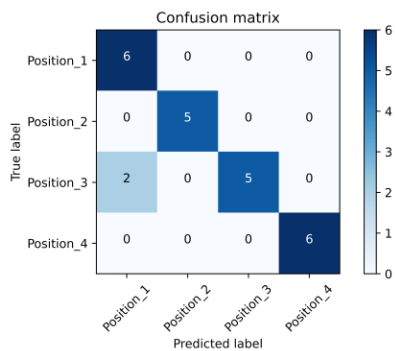

Fig. 9. The confusion matrix of localization in the vehicle by using DT.

Fig. 9 shows that two test samples of human occupancy position 3 are classified into position 1. We also analyzed confusion matrices for other classifiers. Some samples of position 1 and position 3 could be misclassified. Since position 1 is directly in front of position 3, our sitting posture is not strictly fixed during the data collection. Therefore, there will be such a phenomenon of misclassification.

In the classroom scenario, PRF spectra of human signatures collected at 20-gridded locations spaced 1.8 m apart are used as training and test data sets for the GPR regressor. The regression task implemented by GPR is able to demonstrate the capability of the proposed PRF system on precise localization. In the traditional training and test set splitting method, each coordinate location is divided into the training set and test set by 70% and 30%; respectively. Compared with traditional methods, we use 17 out of the 20 gridded locations for the training set and the remaining three locations for the test set. The proposed SHAPR method is more meaningful in localization tasks, as it is beneficial to verify that the movement of human signatures in space produces smooth changes in the PRF spectrum. This is crucial because the localization capability of the GPR regressor for precise coordinates is only possible under the premise that the PRF spectrum changes smoothly. Based on our experience and assumptions, the PRF spectrum will only change very slightly due to the environment when the subject is in the same location. Due to the possible overfitting of the model, the possibility of obtaining high accuracy on the test dataset at the same location is very high, and such results are inaccurate. Therefore, independent and non-interfering data collection for training and testing gridded locations can verify the performance of the GPR model. Moreover, it helps to show GPR model predictions of the subject location at random locations. The predicted locations and actual positions are shown in 0.

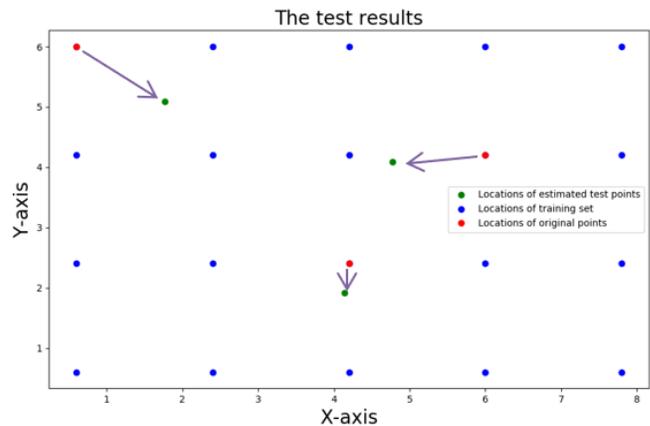

Fig. 10. The localization error in the classroom on the coordinate level by using GPR.

In 0, the blue dots represent the locations of the training set. The red dots represent the original location of the test point. The green dots represent the predicted location based on the spectrum obtained on the blue dots through the GPR model. A set of red and green dots connected by each arrow indicates each set of original and predicted locations. The average residual of all test points after modeling is calculated using the Euclidean distance between green dots and red dots. The range of the average residual is about 0.8 m. When we increase the number of training set points and decrease the number of testing points, the residual will significantly decrease. Therefore, PRF could be utilized to locate a human subject at the coordinate level. Thus, the SHAPR solution can predict the coordinates of an unknown location by GPR based on the PRF and coordinates of the known location and the PRF of the unknown location. Furthermore, if we collect data on more locations for modeling, the location predictions will be better.

D. Activity Recognition

laboratory scenario. In this task, two human subjects are invited to conduct eight activities. Five MLs are utilized to classify eight typical activities, which include using Smartphone, Sitting, Watching TV, Walking, Standing, Exercising, Writing on the Board, and simulating Falls. The activity recognition classification results are shown in Table V.

TABLE V
HUMAN ACTIVITY RECOGNITION ACCURACY FOR FIVE ML MODELS IN THE LABORATORY.

|  | DT | SVM | $k$-NN | RFR | RNN |
| --- | --- | --- | --- | --- | --- |
| **Laboratory** | 85.6% | 81.9% | 99.3% | **99.1%** | 97.0% |

Activity recognition is a particular task because dynamic activities have temporal continuity and location uncertainty and variability compared to static authentication and localization. The subject's dynamic activity continuously affects the PRF spectrum. Considering that the sampling time of each PRF spectral sample is about eight seconds, the dynamic activity of the subjects continuously affects the average power at different frequencies. Since the active tasks are set to tasks such as Walking and Exercise, the location of the subjects is also uncertain and varied. Therefore, dynamic activity recognition continuously affects the average power change over frequency in time and space, which are both a challenge and a feature for activity recognition, because ML algorithms can easily classify walking and standing activities. RFR verifies our conjecture with 99.1% accuracy in classifying activities, and it can easily classify these eight activities. The results of RNN for activity classification are more of our concern due to its ability to handle temporally continuous data. The confusion matrix of using RNN is shown in Fig. 11.

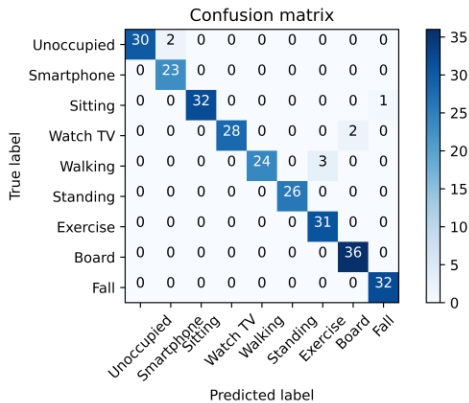

Fig. 11. The confusion matrix of activity recognition in the laboratory by using the RNN.

Fig. 11 shows errors in activity classification, such as the RNN model misclassifying three Walking to Exercise. RNN is originally expected to outperform traditional ML algorithms. RNN models have more weights, biases, and hyperparameters than traditional ML to extract better and fit data features. However, from the experimental results, traditional ML, especially RFR, has higher accuracy than RNN. RNNs generally have a strong ability to deal with continuous data because of their internal Long Short-Term Memory (LSTM) modules which require large data sets to fine-tune the models. For our experiments, the RNN does not perform well in authentication and localization tasks. Possible reasons are that the authentication and localization tasks are static classification tasks and the PRF spectrum is not continuous data. The PRF spectrum has independent average power at each frequency, and there is no specific correlation between each frequency. According to the PRF spectrum in Fig. 3, we notice that the PRF spectrum actually does not change regularly, and only the changes between adjacent frequencies are not very drastic but also there is not overall upward or downward trend. In the dynamic activity recognition task, however, there is a temporal correlation between subjects' activities, which causes the average power corresponding to each frequency in the PRF spectrum to rise or fall regularly. This regular change may be due to subjects approaching or moving away from the SDR antenna at similar speeds, etc., which may regularly affect the PRF spectrum. However, the classification accuracy of RFR is higher than that of RNN, indicating that this regular change has less impact on the overall PRF spectrum, so the ML algorithm of RFR can better process PRF data.

*E. Ablation Study for Multi-Receivers*

Performance varies with the number of receivers. The classification accuracy of the PRF system improves as more receivers are added, as shown in Fig. 12. This is because an increase in the number of receivers results in a larger number of samples, enabling the classifier to recognize more subtle human signature features. Additionally, having receivers with different spatial distributions enables more accurate modeling of human distance behavior. The impact of the number of PRF receivers is similar to that of WiFi access points. However, adding more receivers can lead to several issues, including increased system cost and unnecessary resource consumption. Moreover, deploying receivers optimally, ensuring synchronization among receivers, and adapting the system to the specific scenario are also important considerations. Therefore, achieving a balance between receiver performance and system efficiency remains a crucial area of research.

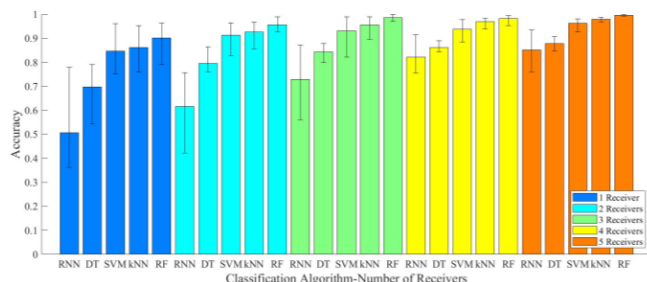

Fig. 12. Comparison of the number of antennas and the accuracy of human activity recognition.

VI. CONCLUSIONS

This paper presents the SHAPR by monitoring the PRF spectrum. Five ML algorithms are performed to realize authentication, localization, and activity recognition. In laboratory and living room scenarios, the human subject's authentication accuracies are 95.6% and 98.7% by using RFR, verifying that different human subjects can generate different signatures on the PRF spectrum. The high accuracy authentication method could be used to authorize people who match the ID database signature. The SHAPR human localization method utilizes a grid level and coordinate level. Human subjects occupying location classification tasks can reach 99.1% accuracy in the vehicle scenario. For the coordinate-level localization, the average error is 0.8 m, which clarified humans at different locations can produce different alterations on the PRF spectrum. In the activity's recognition classification task, eight activities could be classified with 99.1% accuracy by using the RFR, which means different human postures also can generate different signatures on the PRF spectrum. Therefore, our research not only enlarged the applications of the PRF spectrum monitoring, but also found a novel biometric modality, and utilized this modality feature to realize human sensing with comparable accuracy.

Our future work lies in extending PRF technology to crowd-sensing scenarios. Examples of this research include counting multiple human subjects, measuring the speed of the human subject's movement, and tracking the movement of the human subject [43]. Another research direction is the fusion of multimodal passive sensors for more fine-grained human monitoring applications, such as blood pressure detection, respiration reconstruction, gesture recognition, etc. Single-sensor modality has specific limitations, so multi-sensor fusion technology can provide a good solution to achieve recognition and detection of targets. By using the cooperation of radar, lidar, and cameras, human detection and scene recognition can be coordinated with passive recognition technology to improve accuracy and overcome environmental interference.

## References

[1] Y. Feng, S. Yu, H. Peng, Y. -R. Li and J. Zhang, "Detect Faces Efficiently: A Survey and Evaluations," *IEEE Trans. Biom. Behav. Identity Sci.*, vol. 4, no. 1, pp. 1-18, Jan. 2022.

[2] M. Mostofa, S. Mohamadi, J. Dawson and N. M. Nasrabadi, "Deep GAN-Based Cross-Spectral Cross-Resolution Iris Recognition," *IEEE Trans. Biom. Behav. Identity Sci.*, vol. 3, no. 4, pp. 443-463, Oct. 2021.

[3] J. Galbally, S. Marcel and J. Fierrez, "Image Quality Assessment for Fake Biometric Detection: Application to Iris, Fingerprint, and Face Recognition," *IEEE Trans. Image Process.*, vol. 23, no. 2, pp. 710-724, Feb. 2014.

[4] N. Khamsemanan, C. Nattee and N. Jianwattanapaisarn, "Human Identification From Freestyle Walks Using Posture-Based Gait Feature," *IEEE Trans. Inf. Forensics Secur.*, vol. 13, no. 1, pp. 119-128, Jan. 2018.

[5] S. E. Tandogan and H. T. Sencar, "Estimating Uniqueness of I-Vector-Based Representation of Human Voice," *IEEE Trans. Inf. Forensics Secur.*, vol. 16, pp. 3054-3067, 2021.

[6] C. Hao, X. Wan, D. Feng, Z. Feng, and X.-G. Xia, "Satellite-Based Radio Spectrum Monitoring: Architecture, Applications, and Challenges," *IEEE Netw.*, vol. 35, no. 4, pp. 20-27, Aug. 2021.

[7] A. V. Kramarenko, A. V. Kramarenko, and O. Savenko, "A new radio-frequency acoustic method for remote study of liquids," *Sci. Rep.*, vol. 11, no. 1, pp. 1-11, Mar. 2021.

[8] Y. She, T. Tang, G. J. Wen, and H. R. Sun, "Ultra‐high‐frequency radio frequency identification tag antenna applied for human body and water surfaces," *Int. J. RF Microw. Comput.-Aided Eng.*, vol. 29, no. 1, p. e21464, Jan. 2019.

[9] P. E. Watson, I. D. Watson, and R. D. Batt, "Total body water volumes for adult males and females estimated from simple anthropometric measurements," *Am. J. Clin. Nutr.*, vol. 33, no. 1, pp. 27-39, Jan. 1980.

[10] R. F. Kushner and D. A. Schoeller, "Estimation of total body water by bioelectrical impedance analysis," *Am. J. Clin. Nutr.*, vol. 44, no. 3, pp. 417-424, 1986.

[11] U. M. Qureshi et al., "RF path and absorption loss estimation for underwater wireless sensor networks in different water environments," *Sensors*, vol. 16, no. 6, p. 890, Jun. 2016.

[12] L. Yuan and J. Li, "Smart Cushion Based on Pressure Sensor Array for Human Sitting Posture Recognition," *2021 IEEE Sens.*, 2021, pp. 1-4.

[13] H. Xu, M. Wu, P. Li, F. Zhu, and R. Wang, "An RFID indoor positioning algorithm based on support vector regression," *Sensors*, vol. 18, no. 5, p. 1504, May 2018.

[14] J. Liu, H. Mu, A. Vakil, B. Ewing, X. Shen, E. Blasch and J. Li, "Human Occupancy Detection via Passive Cognitive Radio," *Sensors*, vol. 20, no. 15, p. 4248, Jul. 2020.

[15] H. Mu, R. Ewing, E. Blasch, and J. Li, "Human Subject Identification via Passive Spectrum Monitoring," *2021 IEEE Natl. Aerosp. Electron. Conf. (NAECON)*, 2021, pp. 317-322.

[16] H. Mu, J. Liu, R. Ewing, and J. Li, "Human Indoor Positioning via Passive Spectrum Monitoring," *2021 55th Annu. Conf. Inf. Sci. Syst. (CISS)*, 2021, pp. 1-6.

[17] L. Yuan, J. Andrews, H. Mu, A. Vakil, R. Ewing, E. Blasch, and Jia Li, "Interpretable Passive Multi-modal Sensor Fusion for Human Identification and Activity Recognition," *Sensors*, vol. 22, no. 15, p. 5787, Aug. 2022, doi: 10.3390/s22155787.

[18] S. Yang, L. Yuan, and J. Li, " Extraction and Denoising of Human Signature on Radio Frequency Spectrums," *2023 IEEE International Conference on Consumer Electronics (ICCE)*, 2023, pp. 1-6.

[19] D. Roy, T. Mukherjee, M. Chatterjee, E. Blasch, E. Pasiliao, "RFAL: Adversarial Learning for RF Transmitter Identification and Classification," *IEEE Trans. Cogn. Commun. Netw.*, vol. 6, no. 2, pp. 783-801, June 2020.

[20] F. M. Ghannouchi, "Power amplifier and transmitter architectures for software-defined radio systems," *IEEE Circuits Syst. Mag.*, vol. 10, no. 4, pp. 56-63, Nov.2010.

[21] R. Danymol, T. Ajitha and R. Gandhiraj, "Real-time communication system design using RTL-SDR and Raspberry Pi," *2013 Int. Conf. Adv. Comput. Commun. Syst.*, 2013, pp. 1-5.

[22] J. Yaoyao, L. Byunghun, K. Fanpeng, K. Zhaoping, M. Connolly, B. Mahmoudi, M. Ghovanloo, "A Software-Defined Radio Receiver for Wireless Recording from Freely Behaving Animals," *IEEE Trans. Biomed. Circuits Syst.*, vol. 13, no. 6, pp. 1645-1654, Oct. 2019.

[23] T. Chin, Jr., K. Xiong, E. Blasch, "CRAMStrack: Enhanced Nonlinear RSSI Tracking by Using Circular Multi-Sectors," *Journal of Signal Processing Systems*, 93:79-97, 2021. T. Zhang, T. Song, D. Chen, T. Zhang, J. Zhuang, "A Wifi-Based Gesture Recognition System Using Software-Defined Radio," IEEE *Access, vol.* 7, pp. 131102-131113, Sep.2019.

[24] S. Archasantisuk and T. Aoyagi, "The human movement identification using the radio signal strength in WBAN," *2015 9th Int. Symp. Med. Inf. Commun.Technol. (ISMICT)*, 2015, pp. 59-63.

[25] Y. Xu, J. Y. Wang, B. X. Cao and J. Yang, "Multi sensors based ultrasonic human face identification: Experiment and analysis,"*2012 IEEE Int. Conf. Multisens. FusionIntegr. Intell. Syst. (MFI)*, 2012, pp. 257-261.

[26] M. Strohmeier, I. Martinovic and V. Lenders, "A k-NN-Based Localization Approach for Crowdsourced Air Traffic Communication Networks," *IEEE Trans. Aerosp. Electron. Syst.*, vol. 54, no. 3, pp. 1519-1529, Jun. 2018,

[27] C. Dewi and R. -C. Chen, "Human Activity Recognition Based on Evolution of Features Selection and Random Forest," *2019 IEEE International Conference on Systems, Man and Cybernetics (SMC)*, 2019, pp. 2496-2501.

[28] C. Rasmussen and C. Williams, "Gaussian processes in machine learning" *MIT Press*, 2006, Chapter 2, 4-5, 7-8, pp.7-30, 79-128, 151-185.

[29] A. Bekkali, T. Masuo, T. Tominaga, N. Nakamoto and H. Ban, "Gaussian processes for learning-based indoor localization," *2011 IEEE Int. Conf. Signal Process. Commun. Comput. (ICSPCC)*, Xi'an, 2011, pp. 1-6.

[30] L. Yuan, H. Chen, R. Ewing, E. Blasch, and J. Li, "Three Dimensional Indoor Positioning Based on Passive Radio Frequency Signal Strength Distribution," *IEEE Internet Things J.*, Mar. 2023.

[31] S. Yiu and K. Yang, "Gaussian Process Assisted Fingerprinting Localization," *IEEE Internet Things J.*, vol. 3, no. 5, pp. 683-690, Oct. 2016.

[32] B. Ferris, D. Haehnel and D. Fox, "Gaussian processes for signal strength-based location estimation", *Proc. Robot. Sci. Syst.*, Jun.2006.

[33] Y. Liu, X. Wan and X. Sun, "GPU parallel acceleration of target detection in passive radar system," *2016 CIE Int. Conf. Radar (RADAR)*, 2016, pp. 1-4.

[34] J. H. Huang, M. N. Barr, J. L. Garry and G. E. Smith, "Subarray processing for passive radar localization," *2017 IEEE Radar Conf. (RadarConf)*, 2017, pp. 0248-0252.

[35] V. Navrátil, J. L. Garry, A. O'Brien and G. E. Smith, "Utilization of terrestrial navigation signals for passive radar," *2017 IEEE Radar Conference (RadarConf)*, 2017, pp. 0825-0829.

[36] D. Cristallini, I. Pisciottano and H. Kuschel, "Multi-Band Passive Radar Imaging Using Satellite Illumination," *2018 Int. Conf. Radar (RADAR)*, 2018, pp. 1-6.

[37] A. Buffi, P. Nepa and R. Cioni, "SARFID on drone: Drone-based UHF-RFID tag localization," *2017 IEEE Int. Conf. RFID Technol. Appl. (RFID-TA)*, 2017, pp. 40-44.

[38] A. Buffi and P. Nepa, "An RFID-based technique for train localization with passive tags," *2017 IEEE Int. Conf. RFID (RFID)*, 2017, pp. 155-160.

[39] H. He, X. Chen, L. Ukkonen and J. Virkki, "Clothing-Integrated Passive RFID Strain Sensor Platform for Body Movement-Based Controlling," *2019 IEEE Int. Conf. RFID Technol. Appl. (RFID-TA)*, 2019, pp. 236-239.

[40] G. Álvarez-Narciandi, A. Motroni, M. R. Pino, A. Buffi and P. Nepa, "A UHF-RFID gate control system based on a Convolutional Neural Network," *2019 IEEE Int. Conf. RFID Technol. Appl. (RFID-TA)*, 2019, pp. 353-356.

[41] W. Li, W. Zhang and K. W. Tam, "Wireless control for Vision Impaired based on posture effect of UHF RFID cane," *2016 IEEE Int. Conf. RFID Technol. Appl. (RFID-TA)*, 2016, pp. 93-96.

[42] E. Blasch, A. J. Aved, "Physics-Based and Human-derived Information Fusion Video Activity Analysis," *2018 21st Int. Conf. Inf. Fusion (FUSION)*, 2018, pp. 997-1004.

[43] J.M.P. Martinez, R.B. Llavori, M.J.A. Cabo, and T.B. Pedersen, "Integrating Data Warehouses with Web Data: A Survey," *IEEE Trans. Knowledge and Data Eng.*, preprint, 21 Dec. 2007, doi:10.1109/TKDE.2007.190746.

[44] W. Wang, A.X. Liu, M. Shahzad, K. Ling, and S. Lu, "Understanding and modeling of wifi signal based human activity recognition," *Proceedings of the 21st annual international conference on mobile computing and networking*, 2015, pp. 65-76.

[45] F. Zhang, J. Xiong, Z. Chang, J. Ma, and D. Zhang, "Mobi2Sense: empowering wireless sensing with mobility," *Proceedings of the 28th Annual International Conference on Mobile Computing And Networking*, 2022, pp. 268-281.